\documentclass[a4paper,10pt,oneside,onecolumn,titlepage,final]{article}
\usepackage[a4paper,body={6.5in,9in},top=1in,left=1in]{geometry}
\usepackage[lists,nomarkers]{endfloat}
\usepackage{times}
\usepackage{fancyhdr} 
\usepackage{graphicx}
\usepackage{epsfig}
\usepackage{tabularx}
\usepackage{amsmath2000}
\usepackage{amssymb}

\newenvironment{trbitemize1}{\begin{list}{\labelitemi}{\setlength{\leftmargin}{0.5in}}}{\end{list}}

\usepackage[numbers,round]{natbib}

\usepackage{subfigure}

\def\imp#1{\textbf{#1}}
\def\eq#1{Eq.~(\ref{#1})}

\def\yy#1{\relax}

\newcommand{\Var}[1]{\operatorname{Var}[#1]}
\newcommand{\Avg}[1]{\operatorname{E}[#1]}

\newcommand{\vmax}{\ensuremath{v_\text{max}}}

\newcommand{\tRelax}{\ensuremath{t_\text{relax}}}
\newcommand{\NumCars}{\ensuremath{N}}
\newcommand{\RoadLen}{\ensuremath{L}}
\newcommand{\DensMesLen}{\ensuremath{\ell}}

\newcommand{\LocalDens}{\ensuremath{\rho_\ell}}
\newcommand{\densVar}{\ensuremath{\Var{\LocalDens}}} 

\newcommand{\laminar}{laminar}
\newcommand{\jammed}{jammed}

\begin{document}
\begin{titlepage}
\noindent \Large{\textbf{Probabilistic Traffic Flow Breakdown In Stochastic Car
Following Models}}

\vspace{1cm} {\large
\noindent Dominic Jost\\
\noindent Dept.\ of Computer Science, ETH Z\"urich, CH-8092~Z\"urich, Switzerland\\
\noindent Telephone: +41 1 632 2754\\
\noindent Fax: +41 1 632 1374\\
\noindent E-mail: djost@student.ethz.ch}
 
\vspace{0.7cm}{\large
\noindent Kai Nagel\\
\noindent Dept.\ of Computer Science, ETH Z\"urich, CH-8092~Z\"urich, Switzerland\\
\noindent Telephone: +41 1 632 2754\\
\noindent Fax: +41 1 632 1374\\
\noindent E-mail: nagel@inf.ethz.ch} 

\vspace{0.7cm}{\large
\noindent Submission date: August 01, 2002}
 
\vspace{0.7cm}{\large
\noindent Corresponding author: Kai Nagel}
 
\vspace{0.7cm}{\large
\noindent Number of words: 6805}
 
\end{titlepage}

\begin{abstract} 
There is discussion if traffic displays multiple phases (e.g.\ 
laminar, jammed, synchronized) or not.  This paper presents evidence
that a stochastic car following model, by changing one of its
parameters, can be moved from showing two phases (laminar and jammed)
to showing only one phase.  Models with two phases show three states:
two being homogeneous states corresponding to each phase, and a third
state which consists of a mix between the two phases (phase
coexistence).
\\
Although the gas-liquid analogy to traffic models has been widely
discussed, no traffic-related model ever displayed a completely
understood \emph{stochastic} version of that transition.  Having a
stochastic model is however important to understand the potentially
probabilistic nature of the transition.  Most importantly, if indeed
2-phase models describe certain aspects correctly, then this leads to
predictions for breakdown probabilities.  Alternatively, if 1-phase
models describe these aspects better, then there is no breakdown.
Interestingly, such 1-phase models can still allow for jam formation
on small scales, which may give the impression of having a 2-phase
dynamics.
\end{abstract}

\section{INTRODUCTION}

Both from an operations and from a design perspective, the capacity of
a road is an important quantity.  Clearly, if demand exceeds capacity,
queues will form, which represent a cost to the driver and thus to the
economic system.  In addition, such queues may impact other parts of
the system, for example by spilling back into links used by drivers
who are on a path that is not overloaded.  

For a variety of reasons, however, capacity is not a deterministic
fixed quantity.  It is possible that one day a queue forms and the
next day not, and this may even happen in spite of demand being larger
on the second day.  In consequence, any definition of capacity needs
to take its stochastic nature into account.

For example, one could measure flow in 15-min intervals, say from 6am
to 6:15am, from 6:15am to 6:30am, etc.  One could then take the daily
maximum of these values, and average the result over many typical
workdays.

As an alternative, one could measure flow as a function of density
irrespective of any other state variables.  One could
then obtain the average flow for each density interval, and the
maximum of these flow-values would represent capacity.

All of these measurements have the property that they result in an
expected value, i.e.\ in a number that, for a given day, can be
exceeded or not be reached.    
%
In consquence, it is useful to develop models of traffic which reflect
the stochastic nature of traffic.  Clearly, the stochasticity can come
at many different levels: demand can vary; road conditions can vary;
driving behavior can vary; etc.  These different contributions to
stochasticity will have different influences, which need to be
debated.
In this paper, we want to concentrate on road capacity. 
We understand that there is active research to eventually include
aspects of stochastic transitions into the Highway Capacity
Manual~\cite{Lily:personal}.

The starting point for our work are simple single-lane car following
models.  These models are typically either of the type $v(t+\tau_v) =
f(g,\Delta v, ...)$ or of the type $a(t + \tau_a) = h(g,\Delta v,
...)$, where $v(t)$ is the velocity of a car at time $t$, $\Delta v$
is the velocity difference to the car ahead, and $a$ is the
acceleration.  $g$ is the gap to the car ahead, where $g = \Delta x -
\ell$, with $\Delta x$ the front-buffer-to-front-buffer distance, and
$\ell$ is the space the car occupies in a jam.  These models can for
example be found in~\cite{Newell:1961}
\[
v(t\!+\!\tau_v) = V(g(t)) 
\hbox{\ \ with\ \  }
V(g) = v_f - v_f \, \exp( - \lambda g / v_f ) \ , 
\label{eq:Newell}
\]
in~\cite{Bando94}
\[
a(t) = \alpha \cdot \Big( V(g(t)) - v(t) \Big)\ ,
\hbox{\ \ with\ \ }
V(g) = v_f \cdot (\tanh(g+\ell) - \tanh(\ell)) \ , 
\label{eq:Bando}
\]
or in~\cite{theGang}
\[
a(t\!+\!\tau_a) \propto \frac{[v(t\!+\!\tau_a)]^l}{[\Delta x(t)]^m} \Delta v(t) \ .
\label{eq:theGang}
\]
Additional parameters here are $v_f$ (the free speed), $\lambda$, $l$,
and $m$.

When these models are implemented on a computer, they need to be
discretized in time, and one has to concern oneself with the size of
the integration time step, $\Delta t$.  A typical discretization is
\[
a(t) = \hbox{given by the model}
\]
\[
v(t) = v(t - \Delta t) + \Delta t \, a(t) \ ,
\]
and
\[
x(t + \Delta t) 
= x(t) + \Delta t \, v(t) \ .
\]
These discretizations are meant to approach the original coupled
differential equations for $\Delta t \to 0$, and there is a whole body
of literature available for this.  \yy{citation} Once time delays (via
$\tau > 0$) are introduced into such equations, numerical treatment
becomes considerably more difficult, and much less is known about
efficient numerical integration. \yy{citation}

Given that state of affairs, it makes sense to look for computational
models which are not based on the limit $\Delta t \to 0$, but which
generate useful results also for relatively large time steps of, say,
one second.  The model that we will use in this paper has been
introduced by Krauss~\cite{Krauss:phd}; it is a variant of a model
used by Gipps~\cite{Gipps:following}.  The Krauss model has been shown to be
free of collisions (i.e.\ that $g < 0$ never occurs).

In addition to being crash free at large time steps, the Krauss model
is also stochastic.  The important parameter for our study is a noise
amplitude $\epsilon$, which we will vary from 0.5 to 2.  For $\epsilon
<0$ or $\epsilon \ge 2$ the model leaves the range of where it is
plausible for traffic.

Our main results are the following:
\begin{trbitemize1}

\item For medium $\epsilon$, there are three states of traffic, which
we will call laminar, mixed, and jammed.  The state depends on the
density.  Laminar, occuring at low density, means that there are no
stopped cars; jammed, occuring at high density, means that no car is
driving at full speed; and mixed, occuring at intermediate density,
means that the system is a mix of laminar and jammed traffic.  In the
mixed state, traffic is strongly inhomogeneous; the mixed state is
often called the coexistence state.

It is important to note that there are three states (laminar, jammed,
mixed) but only two phases (laminar, jammed).  The phases refer to
homogeneous sections of the system; the state refers to the system as
a whole.

\item For large $\epsilon$, there is only one phase of traffic and
therefore only one state.  When going from low to high density, cars
move closer and closer together, but traffic remains homogeneous at
all times.

\item At some $\epsilon$ in between, there is a transition from the
1-phase to the 2-phase regime.  

\item In the Krauss model, changes of $\epsilon$ also change the
average acceleration.  This is an unfortunate coincidence, and we
believe that our general results regarding the number of phases are
not related to this effect.

\item Deterministic models, formulated either as car following models
or as fluid-dynamical models, can display 1-phase or 2-phase behavior.
They can however \emph{not} display stochastic transitions between the
phases.

\end{trbitemize1}

The results are important for model building as well as for
understanding field measurements.  In a 2-phase model, theory predicts
that there can be a hysteretic transition from the laminar to the
mixed state \emph{without a change in density}.  This means that, at a
given density, traffic can operate in the laminar flow state for long
times, until it will eventually ``break down'' and switch to the mixed
state.  In a 1-phase model, this is impossible, and there is only one
phase for any given density.  

A direct consequence of this is that, if traffic follows a 1-phase
model, any initial jam will ``smear out'' and thus eventually go away,
\emph{even with unchanged traffic conditions}.  Conversely, in a
2-phase model, jams ``pull themselves together'' and thus stabilize
themselves.  

This paper starts with Sec.~\ref{sec:phases} which describes the
general idea of a gas-liquid transition.  Sec.~\ref{sec:simulations}
describes the general simulation setup including the car following
model that is used, discusses space-time plots of the resulting
dynamics, and investigates transients vs.\ the steady state.
Sec.~\ref{sec:phase-diag} then establishes how a mixed state can be
nuemrically detected for a given model.  Sec.~\ref{sec:det} discusses
how these results relate to deterministic model; the paper is
concluded by a discussion and a summary.

\section{PHASES IN TRAFFIC}
\label{sec:phases}

At a first glance, our results may seem largely irrelevant to traffic
operations.  As already discussed above, we contend that it is not.
There are thoughts about extending the Highway Capacity Manual so that
it includes the concept of stochastic traffic
breakdown~\cite{Lily:personal}.  This could for example mean that, for
certain flow levels, one would include a curve describing the
probability that traffic flow has not broken down as a function of
time.  In order for such a description to make sense, the existence of
spontaneous flow break-down needs to be established.  Some researchers
however have doubts that flow breakdown truly
happens~\cite{Cassidy:no-2-phase,Mauch:Cassidy:no-2-phase}.  This
paper will contribute to the theoretical understanding of the issue,
by showing how one model can move from displaying breakdown to not
displaying breakdown.

In fact, the analogy between a gas-liquid transition and the
laminar-jammed transition of traffic was pointed out many times
(e.g.~\cite{Montroll,Bando94}).  \yy{need to look up real sugiyama
reference} The description of traffic in the well-known
2-fluid-model~\cite{2fluid:basis} implicitely assumes the existence of
two phases; and all simulation models which use spatial queues
(e.g.~\cite{DYNAMIT,DYNASMART,Gawron:simple}) will display two phases
because of the definition of the dynamics.  The two phases in models
with queues are however much easier to understand than the phases in
more realistic models.

\begin{figure}[t]
\centerline{%
\subfigure[]{
\begin{minipage}[c]{0.5\hsize}
\includegraphics[width=\hsize]{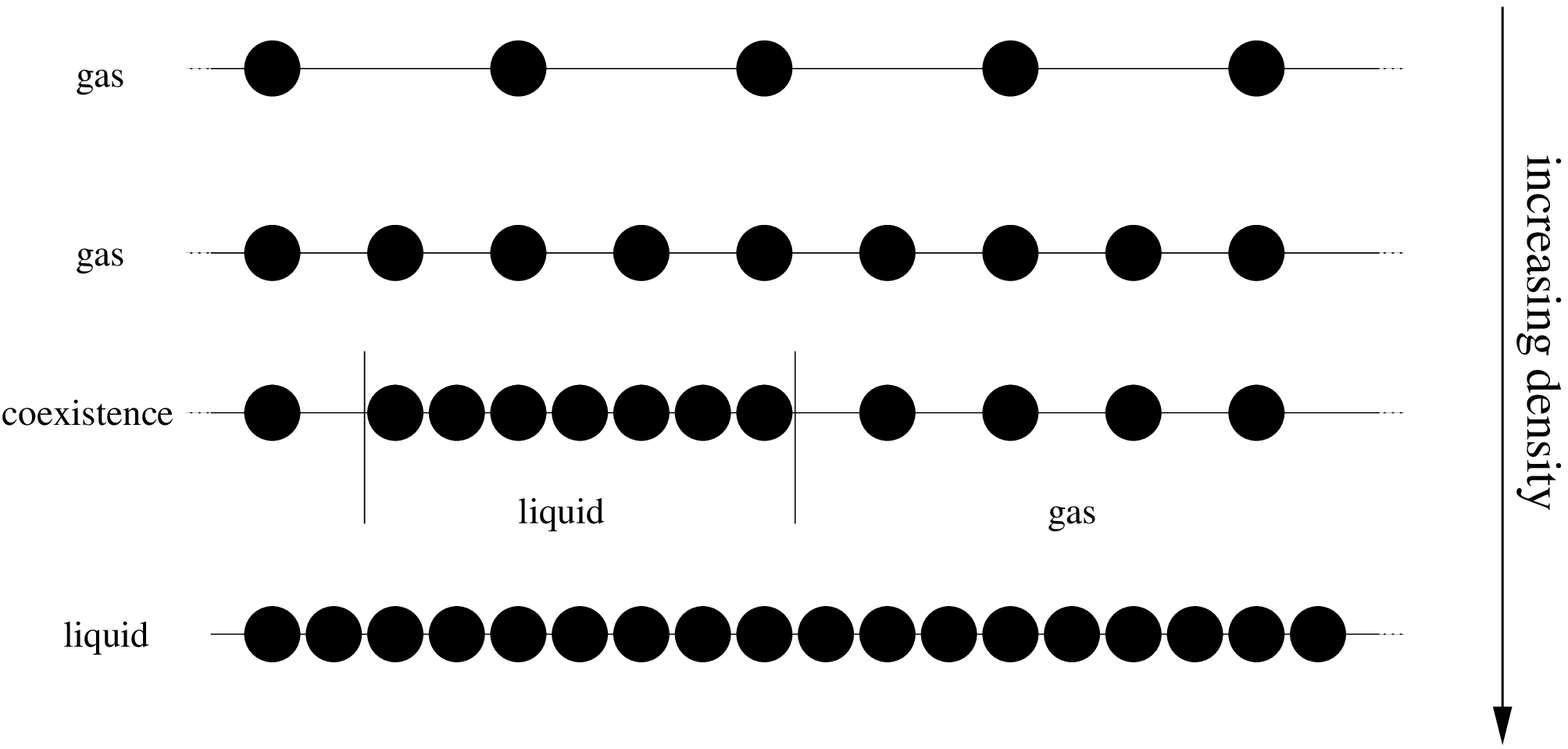}
\end{minipage}
\label{fig:gasLiquid}
}
\subfigure[]{
\begin{minipage}[c]{0.5\hsize}
\includegraphics[width=\hsize]{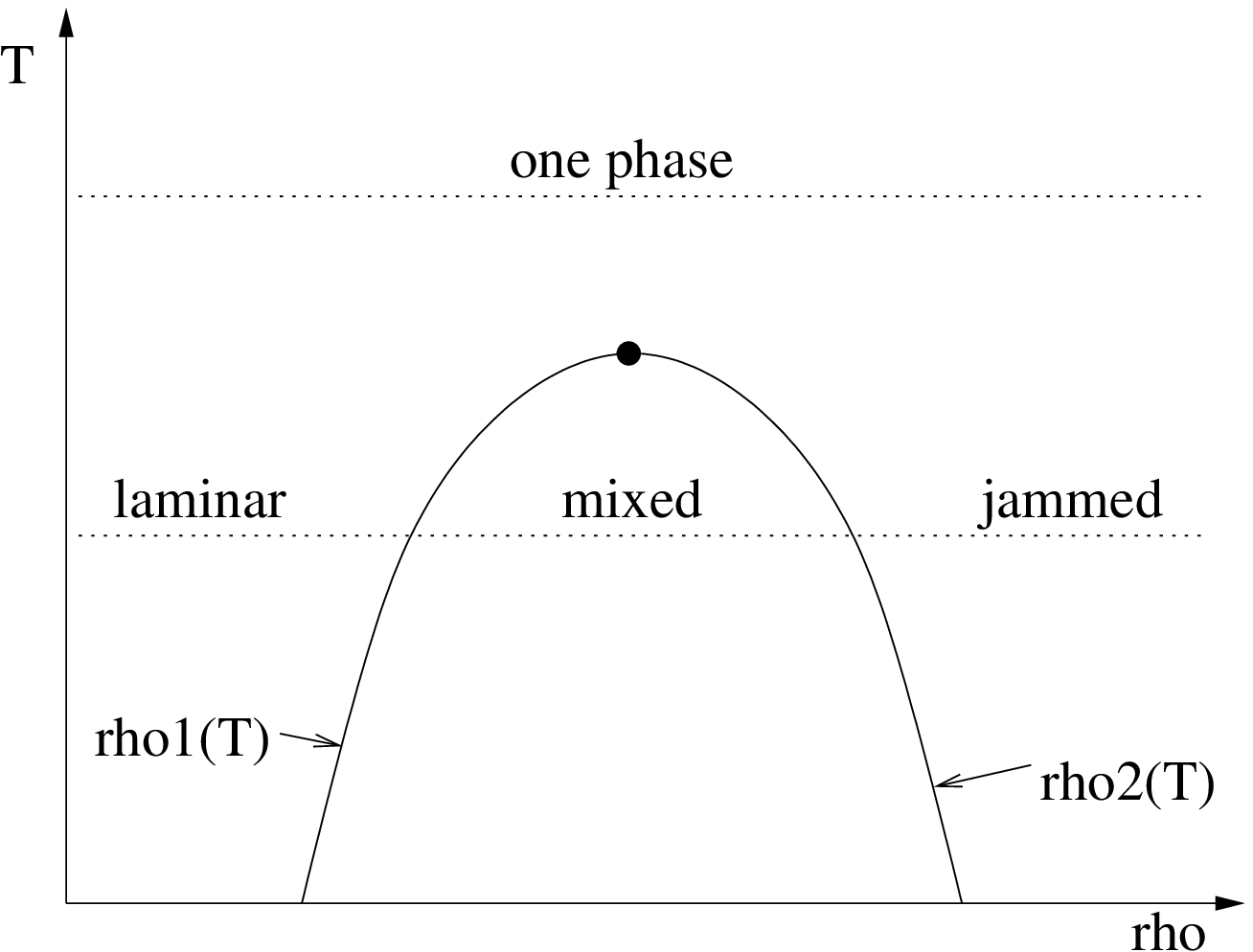}
\end{minipage}
\label{fig:RhoTemp}
}
}
\subfigure[]{
\begin{minipage}[c]{\hsize}
\begin{center}
  $\begin{array}{c@{\hspace{0.5cm}}c}
    \epsfig{file=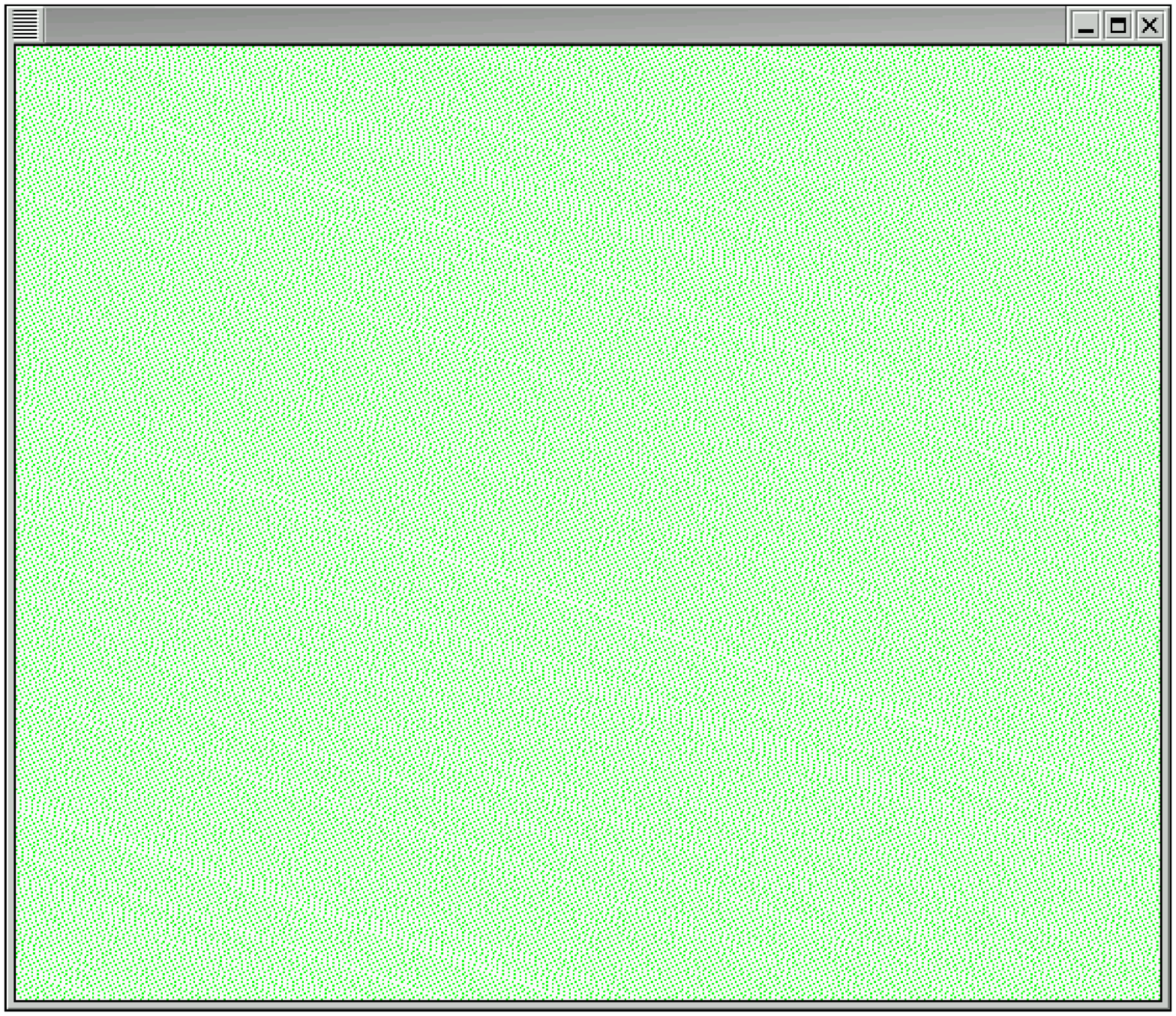, scale=0.25} &
    \epsfig{file=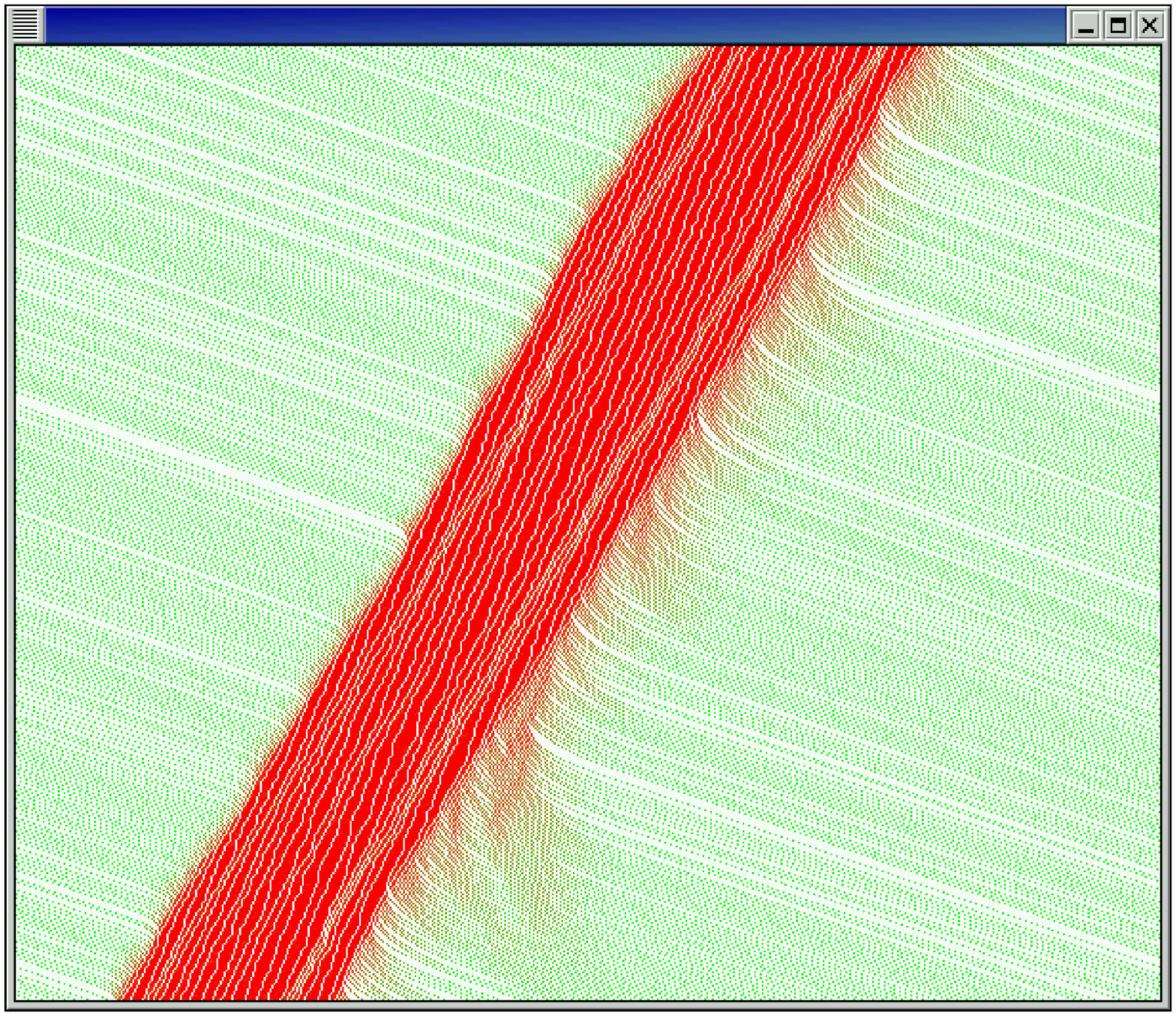, scale=0.25} \\
    \textbf{(i)} \epsilon=0.9,\: \rho=0.2 & \textbf{(ii)} \epsilon=1.0,\: \rho=0.3 \\
    \\
    \epsfig{file=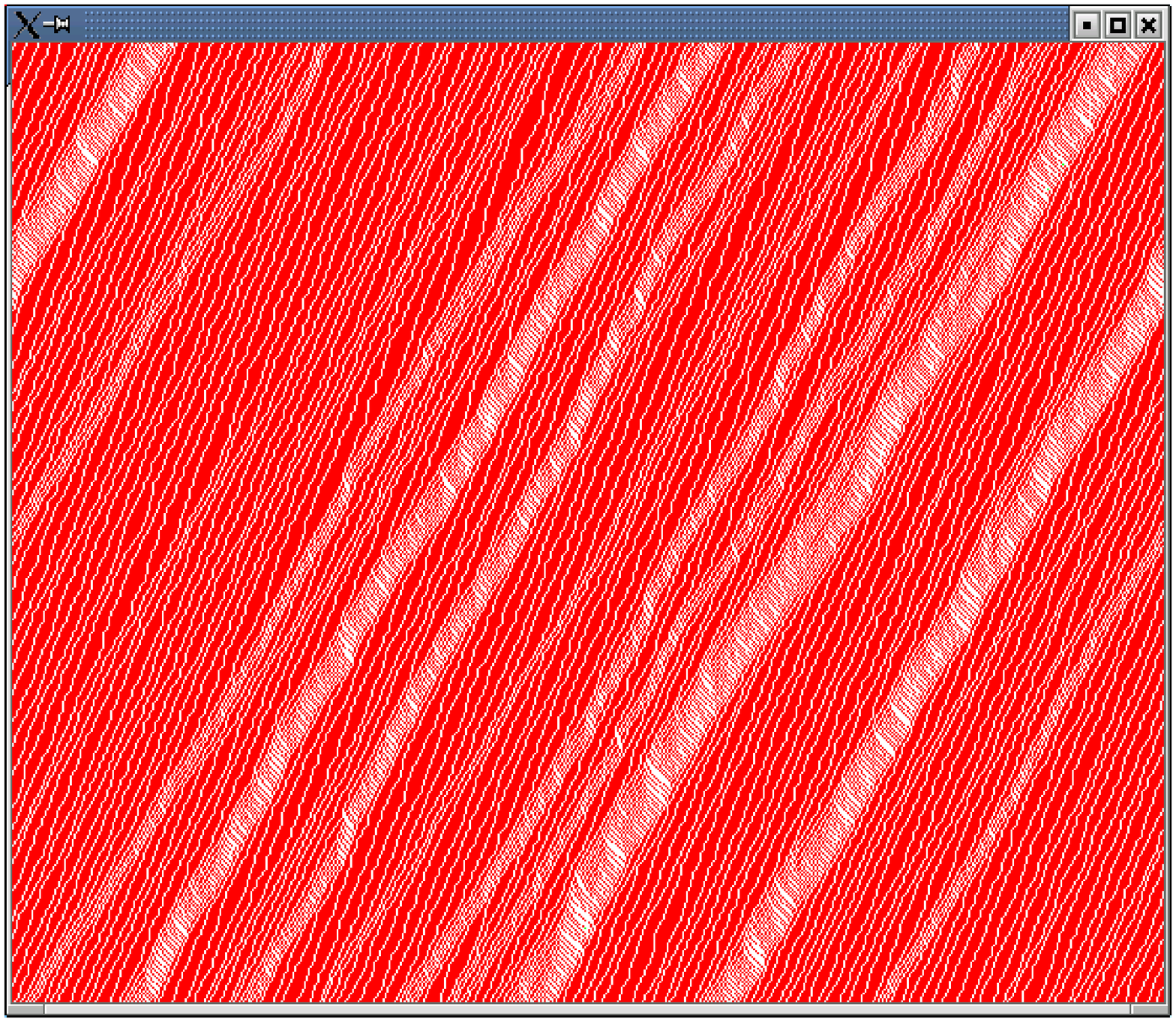, scale=0.25} &
    \epsfig{file=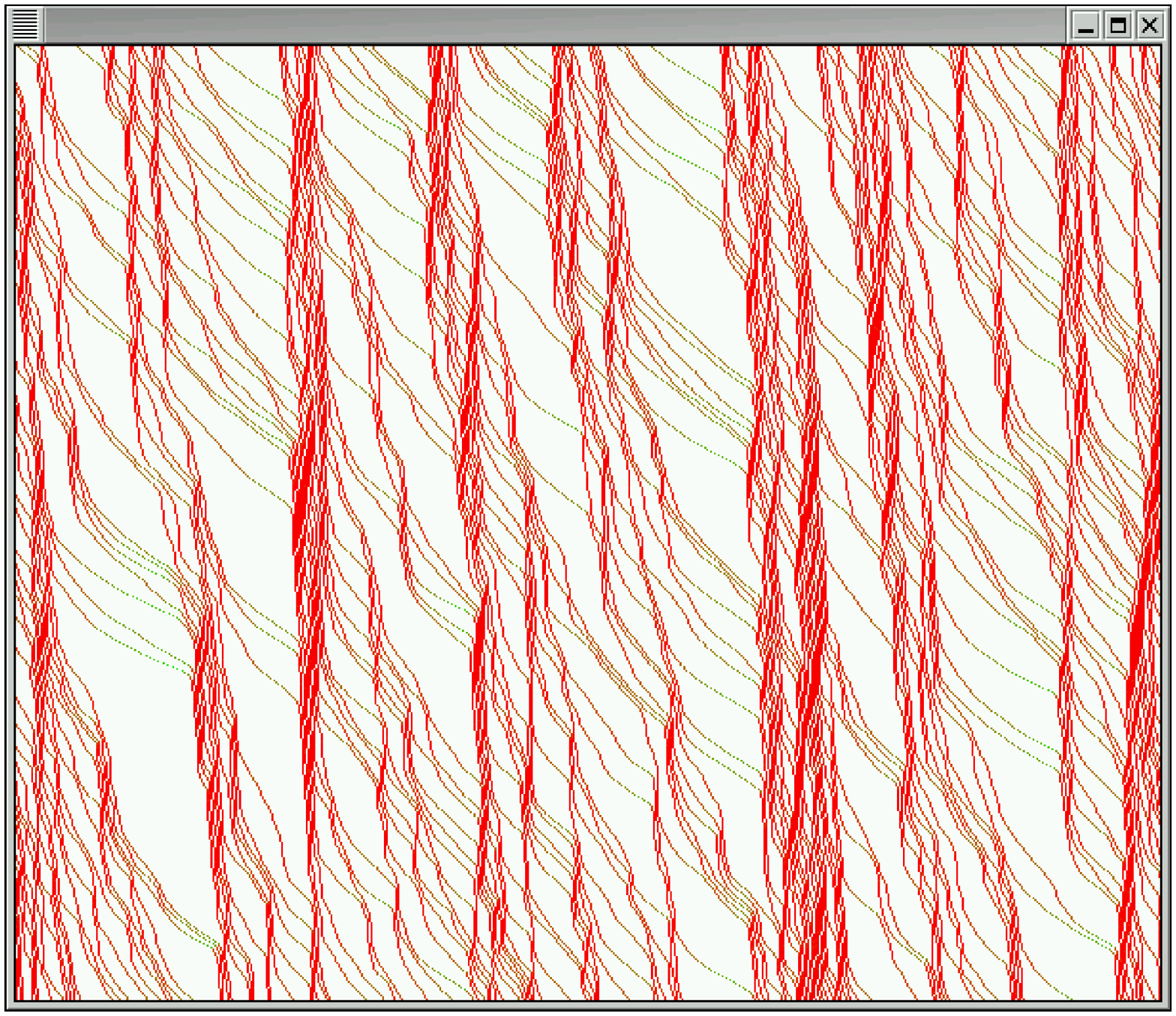, scale=0.25} \\
    \textbf{(iii)} \epsilon=1.0,\: \rho=0.75 & \textbf{(iv)} \epsilon=1.8,\: \rho=0.2
\end{array}$
\end{center}
\end{minipage}
\label{KraussPics}
}
\begin{center}
\caption{\bf %
(a) Schematic representation of the gas-liquid transition in one
dimension.  (b) State of the gas-fluid model as a function of
the density and the temperature.
(c) Space-time plots for different parameters. Space
  is horizontal; time increases downward; each line is a snapshot;
  vehicles move from left to right; fast cars are green, slow cars
  red. $\RoadLen=600$ for all plots.%
}
\end{center}
\end{figure}

In a gas-liquid transition, one observes the following (see also
Fig.~\ref{fig:gasLiquid}~left):
\begin{trbitemize1}

\item In the gas phase at low densities, particles are spread out
throughout the system.  Distances between particles vary, but the
probability of having two particles close to each other is very small.

\yy{can we make this more precise??}

\item In the liquid phase at high densities, particles are close to each
other.  There is no crystalline structure as in solids, but the
density is similar and in some cases (e.g.\ in water) even high in the
liquid than in the solid phase.  Because of the fact that the
particles are so close to each other, it is difficult to compress the
fluid any further.

\item In between, there is the so-called coexistence state, where gas
and liquid coexist.  In typical experiments in gravity, the liquid
will be at the bottom and the gas will be above it.  Without gravity,
as well as for example within clouds, droplets form within the gas and
remain interdispersed.  In clouds, small droplets will eventually
merge together into bigger droplets, which will fall out of the cloud
as rain.  Without gravity, the droplets will just merge but never fall
out.  The final state of the system is having one big droplet of
liquid, surrounded by gas.

\end{trbitemize1}

The kinetics of the droplet formation (e.g.~\cite{Lifshitz:etc:book})
is ruled by a balance between surface tension and vapor pressure.
\yy{Dampfdruck??}  Since surface tension pulls the droplet together,
it has a tendency to push particles out of the droplet.  Vapor
pressure is the balancing force -- it pushes particles into the
droplet.

Surface tension depends on the droplet radius -- the smaller
the droplet, the larger the surface tension.  The result is that, when
coming from small densities, there is a regime where large droplets
would already be stable, but small droplets are not.  In equilibrium,
this density would be in the coexistence state, but coming from low
density, the homogeneous gaseous phase can survive for some time.
This way, meta-stability can be explained in the gas-liquid
transition, and by extension also in traffic models.  

When compressing the system beyond this point, at density $\rho_1$,
the following can be observed: Pressure will \emph{not} go up any
further.  Instead, more droplets will form, and the density outside
the droplets will remain constant.  Pressure will not go up until all
the space is used up by droplets, i.e.\ all particles are moved into
the liquid phase.  Only at this point, at density $\rho_2$, will
pressure go up again when compressing the system.

The state between $\rho_1$ and $\rho_2$ is called
\imp{coexistence state}. In it, gas and liquid coexist, and changes on the
global density are absorbed by the fraction of space and particles
that the liquid occupies.  The interface between droplets and gas
remains the same no matter what the global density is.

In addition to the coexistence, when keeping density constant one will
observe coagulation as already mentioned earlier: The many original
small droplets will coagulate to fewer and fewer large droplets; in
the limit there will only be one large droplet.

A direct consequence of the above behavior is hysteresis:
\begin{trbitemize1}

\item When coming from low densities, it is possible to go beyond
$\rho_1$ and still remain in the gaseous phase.  Only after some
waiting time will, by a fluctuation, some particles get close enough
to each other to start the formation of a droplet.

\yy{check.  In particular, how do we reach relatively stable
supercritical states?  Seems that in certain Krauss models, particles
repulse each oterh???}

\item When coming from densities above $\rho_1$, it is possible that
the droplet survives for some time even below $\rho_1$.  

\end{trbitemize1}

When increasing temperature with the above model, the 2-phase
structure will eventually go away.  This happens via $\rho_1$ and
$\rho_2$ approaching each other and eventually meeting.  That is,
depending on the temperature, a fluid system will either display
transitions from gas to coexistence and from coexistence to liquid, or
there will be \emph{no transition at all} (Fig.~\ref{fig:gasLiquid}~right).

We will now move on to describe the supporting evidence for our
claims.  As is typical in computational statistical physics, our
evidence is based on computer simulations.  It is however backed up by
generic knowledge about phase transitions as they are well understood
in physics.
\section{THE SIMULATIONS}
\label{sec:simulations}

\subsection{Krauss Model}

The velocity update of the Krauss
model~\cite{Krauss:etc:continuous,Krauss:thesis} reads as follows:
\begin{eqnarray}
v_{\rm safe} & = & \tilde v(t) +  
\frac{g(t) -  \tilde v(t)\tau}{\overline{v}/b + \tau}
\label{eq:SK_vSafe} 
\\
v_{\rm des} & = & \min \{ v(t) + a \, \Delta t, v_{\rm safe} , v_{\rm max} \}
\label{eq:SK_vDet} \\
v(t+\Delta t) & = & \max \{ 0 , v_{\rm des} - \epsilon \, a \, \eta \}
\label{eq:SK_updateV} \ .
\end{eqnarray}
$\tilde v$ is the speed of the car in front, $\overline v = (v +
\tilde v)/2$ is the average velocity of the two cars involved, $\vmax$
is the maximum allowed velocity, $a$ is the maximum acceleration of
the vehicles, $b$ their maximum deceleration for
$\epsilon=0$,
$\epsilon$ is the noise amplitude, and $\eta$ is a random number in
$[0,1]$.  The meaning of the terms is as follows:
\begin{trbitemize1}

\item \eq{eq:SK_vSafe}: Calculation of a ``safe'' velocity.  This is
the maximum velocity that the follower can drive when she wants to be
sure to avoid a crash.  The main assumption is that the car ahead will
never decelerate faster than $b$, and that the car of the follower can
also decelerate with up to $b$.

\item \eq{eq:SK_vDet}: The desired velocity is the minimum of: (a)
current velocity plus acceleration, (b) safe velocity, (c) maximum
velocity (e.g.\ speed limit).

\item \eq{eq:SK_updateV}: Some randomness is added to the desired
velocity.  

\end{trbitemize1}
After the velocities of all vehicles are updated, all vehicles
are moved.

The deterministic limit $\epsilon=0$ of the Krauss model has been
proven to be free of crashes for numerical time steps $\Delta t$
smaller than or equal to the reaction time, $\tau$.  We will use
$\Delta t = \tau = 1$ as has conventionally been used for the Krauss
model.  Even for $\epsilon>0$, as was used for our studies, we never
observed vehicles getting closer than their minimum distance.  We
further use $a=0.2$, $b=0.6$, $\vmax=3$ for all simulations.

The model is free of units; this is a property that it has inherited
from the cell-based cellular automata models.  A reasonable
calibration is: time steps correspond to seconds and cells correspond
to $7.5$~meters.  The reaction time then is assumed to be $1$~sec, and
$\vmax=3$ corresponds to $22.5$ m/s or $81$~km/h.  $a=0.2$ corresponds
to a maximum acceleration of $1.5$ m/s per second or $5.4$ km/h per
second. $b=0.6$ corresponds to a maximum deceleration of $16.2$~km/h
per second.

All simulations are done in a 1-lane system of length
\RoadLen\ with periodic boundary conditions (i.e. the road is bent
into a ring).  Let $\NumCars$ be the number of cars on the road. The
(global) density is $\rho = \frac{\RoadLen}{\NumCars}$.

\subsection{Pictures}

Before analysing the Krauss-model numerically, it is instructive to
look at the space-time plots in Fig.~\ref{KraussPics}.  Space-time
plots are pictures of the time evolution of the system.  In
Fig.~\ref{KraussPics}, vehicles drive to the right and time goes
down.  Each row of pixels is a ``snapshot'' of the state of the road.
In principle, one can reconstruct the trajectory of a particular car
by connecting the corresponding pixels.  At the displayed resolution
this is however close to impossible and it is mostly the larger scale
traffic jam structure that one observes.  Traffic jams move
\emph{against} the direction of driving. 
\begin{trbitemize1}
\item [i)]
The laminar state: All cars drive at
  high speed. The available space is shared evenly among the cars. The
  traffic is very homogeneous.
\item [ii)] The mixed state: The slow cars are all together in
  one big jam. On the rest of the road, the cars drive at high speed.
  The traffic is very inhomogeneous.
\item [iii)] The jammed state: The density is so high that not a
  single car can drive fast. As in a), the traffic is very
  homogeneous.
\item [iv)] The single phase at high $\epsilon$: Many small jams
  are distributed over the whole system. There is neither a larger
  area of free flow, nor a major jam. The traffic is homogeneous.
  
  Note that ``homogeneous'' here means ``homogeneous on large
  scales''.  What this means is that there is a spatial measurement
  length $\ell$ above which all density measurements return the same
  value.\footnote{%
More precisely, the fluctuations from one measurement to the next are
the same as in case (i).
} If a system goes from a 2-phase to a 1-phase model, then even in the
regime which technically has only one phase, structure formation on
small scales is still possible.  Fig.~\ref{KraussPics} bottom right is
indeed an example for this.  Only with larger distance from the
2-phase model will these structures go away.

\end{trbitemize1}

\subsection{Defining a jam}

In order to make progress, one needs to define where a jam starts and
where it ends.  Our definition of homogeneity (see later) will not
depend on this, however.  A \emph{jam} is a sequence of cars
driving with speed less or equal $\vmax / 2$. The cars between
two neighbouring jams are in \emph{laminar} flow.

This definition is very simple, but will not always correspond to our
natural understanding of the word jam. Thus, whether a car is jammed
or not according to this definition is just a starting point and not
the final answer.

\subsection{Initial Condition And Relaxation}

For many parameters of the Krauss model, there is a unique equilibrium
state, which the system will attain after a finite time \tRelax, no
matter how it was started. However, deciding when the equilibrium is
reached is not trivial (running the simulation for $t \rightarrow
\infty$ clearly is not an option).

Let $r_t$ be the state of the road at time $t$. To find the
equilibrium value of some property, $E[f(r_{\tRelax})]$, we use the
following idea: For small $t$, $E[f(r_t)]$ will depend on the initial
condition. With increasing time, $E[f(r_t)]$ converges towards the
equilibrium value. Assume the convergence is from above. Now we need
another initial condition that approaches the equilibrium value from
below. Once these two sequences are close enough together, the
equilibrium value is found. Unfortunately, it cannot be guaranteed
that the value thus obtained really is the equilibrium value.

We use the following two initial conditions:
\begin{trbitemize1}
\item \laminar: The cars are positioned equidistant over the road
  with speed zero.
\item \jammed: all cars are cramped together in a big jam without any
  gap. Their speed is zero.
\end{trbitemize1}

\begin{figure}[t]
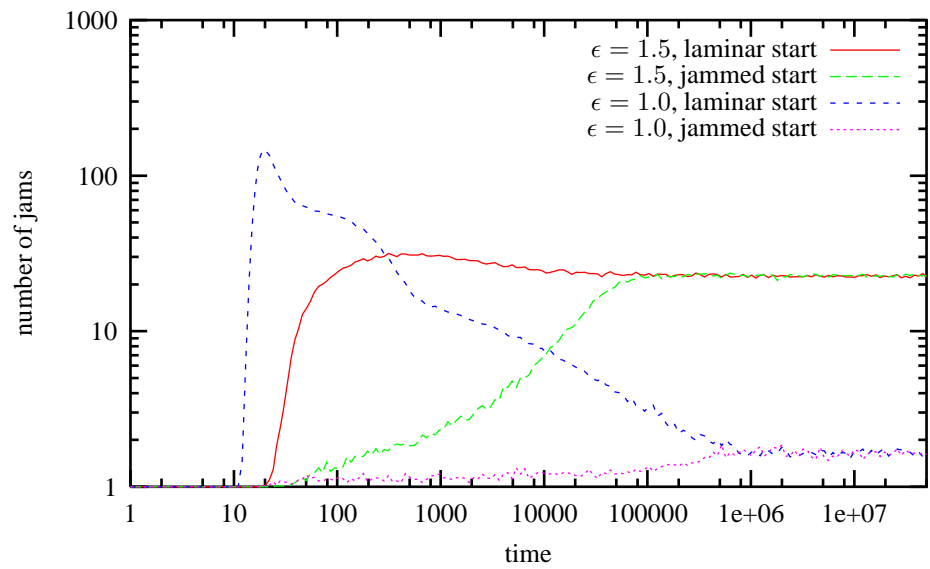

\input KraussNumJamEvol
\label{KraussNumJamEvol}
\caption{\bf Time evolution of the number of jams. All four curves are for 1000
cars.  Each curve is an average over many realizations, each with a
different random seed. \yy{how is this averaged}}
\end{figure}

An example of this method is shown in Fig.~\ref{KraussNumJamEvol}.
For $f(.)$ the number of jams was used.  Since both initial conditions
start with $v=0$, the criterion finds one large jam.  Vehicles then
accelerate, but because of interaction will form small jams.  For that
reason, the laminar start leads to many jams very quickly.  From then
on, the number of jams goes down, because jams coagulate.  In
contrast, when starting with a large single jam, than that jam remains
the only one in the system for large times.  In
Fig.~\ref{KraussNumJamEvol}, we see that for $\epsilon=1.0$, the
system eventually goes to a state where it has, in the average, about
1.8 jams.  In contrast, with $\epsilon=1.5$, the system converges to
an average of more than 20 jams.  Also, the figure shows that the
system goes to those long-run states no matter how it starts.


\begin{figure}[t]
\centerline{%
\begin{minipage}[c]{0.8\hsize}
\input{KraussDensVar.tex} 
\end{minipage}
}
\centerline{%
\begin{minipage}[c]{0.8\hsize}
\input{KraussDensVarCont.tex}
\end{minipage}
}
\caption{\bf %
3d-plot and isolines of the density variance. The outermost isoline is
$\densVar=0.01$, the innermost $\densVar=0.09$. $\RoadLen=1000$ and
$\DensMesLen=62.5$
}
\label{fig:KraussDensVar}
\end{figure}

\section{ESTABLISHMENT OF A PHASE DIAGRAM VIA A MEASURE OF INHOMOGENEITY}
\label{sec:phase-diag}

In this section, a criterion is established that distinguishes
homogeneous from mixed states.  As pointed out before, mixed states
are characterized by the coexistence of laminar and jammed traffic.
As we have also said, if we wait long enough, the phases will
coagulate, leading to exactly one laminar and one jammed section in
the system.   When approaching the boundaries of the mixed regime,
this characterization will become less clear-cut, and it may be
possible to have more than one jam.  Typically, there will be one
major jam and many small ones, but for many measurement criteria this
will cause enough problems to no longer be able to differentiate
between the mixed and a homogeneous state.  Our criterion will only
show a gradual decrease in differentiating power.

As already discussed earlier, it should be noted that some states that
are called ``homogeneous'' in this paper may appear inhomogeneous to
an observer.  An example for this is Fig.~\ref{KraussPics} bottom
right.  As said before, these states are ``homogeneous on large
scales'', which is the important criterion here.  Essentially, this
means that for system size $L \to \infty$ and measurement interval
$\ell \to \infty$ (but $\ell \ll L$), all density measurements will
eventually return the same value.  This will not be the case for
``mixed'' states.

The criterion is defined as follows: Partition the road into segments
of length \DensMesLen\ (for simplicity let \DensMesLen\ divide
\RoadLen\ without remainder). For each segment the local density
\LocalDens\ can be computed as the number of cars in that segment
divided by \DensMesLen. An interesting value is the variance of the
local density:
\begin{equation}
\densVar = \frac{1}{\RoadLen/\DensMesLen}
\sum_{i=1}^{\RoadLen/\DensMesLen} (\LocalDens(i)  -
\Avg{\LocalDens})^2 \ ,
\end{equation}
where $E[.]$ is the expected value, which in our case is the same as
the systemwide density.  Note that since the density lies within
$[0,1]$, the variance cannot exceed $1/4$.

What this value picks up is how much each individual measurement
segment of length $\ell$ deviates, in terms of its density, from the
average density.  Assume a system consisting of jammed and laminar
traffic.  If there is a jam in one segment, then the segment's density
will be much higher than the average density.  Conversely, if there is
only laminar traffic in a segment, then the segment's density will be
much lower than the average density.  $\densVar$ takes the average
over the square of these deviations.




\begin{figure}
\end{figure}


Fig.~\ref{fig:KraussDensVar} shows this value as a function of the
global density $\rho$ and the noise parameter $\epsilon$.  Look at it
for fixed $\epsilon$, say $\epsilon = 1$.  One sees that at densities
up to $\rho \approx 0.2$, the value of $\densVar$ is close to zero,
indicating a homogeneous state, which is in this case the laminar
state.  Similarly, for densities above $\rho \approx 0.8$, $\densVar$
is again close to zero, indicating a homogeneous state, which is in
this case the jammed state.  In between, for $0.2 \lesssim \rho
\lesssim 0.8$, the value of $\densVar$ is significantly larger than
zero, indicating a mixed state.  

Now slowly increase $\epsilon$.  We see that the laminar regime ends
at smaller and smaller densities, while the jammed regime starts at
smaller and smaller densities.  The latter means that for large
$\epsilon$, the jammed phase has many relatively small holes, which
reduce the density, but do not break the jam.  At $\epsilon \approx
1.7$, the mixed phase completely goes away; for larger $\epsilon$, we
do not pick up any inhomogeneity at \emph{any} density.  Compare this
to the theoretical expectation in Fig.~\ref{fig:RhoTemp}, where for
increasing $T$ the two densities eventually merge and thus the
different phases go away.  Note that close to the transition the
system still looks like it possesses different phases (see
Fig.~\ref{KraussPics}(iv) and locate the corresponding $\epsilon=1.8$
and $\rho=0.2$ in Fig.~\ref{fig:KraussDensVar}).  These structures do
however exist \emph{on small scales only}; when averaging over larger
segments, then all segments contain exactly the same density.  A
segment length of $\ell=62.5$, as used in the figure, is already
sufficient in order to not measure any inhomogeneity for the state in
Fig.~\ref{KraussPics}(iv).

Remember again that $\epsilon$ is a model parameter while $\rho$ is
a traffic state.  That is, once one has settled for an $\epsilon$, the
model behavior is fixed, and one has decided if one can encounter a
second phase or not.  \emph{If} one can encounter a second phase, it
will come into existence through changing traffic demand throughout
the day -- traffic can move from the laminar into the mixed and
potentially into the jammed state and back.

\yy{note for later: this has the curious consequence that also a
bottleneck queue can display phase separation.  hoho}

As a side remark, let us note that there is also another 1-phase
regime for $\epsilon \to 0$.  Albeit potentially interesting, this is
outside the scope of this paper.

\yy{there is stuff here that I commented out because I was not sure if
I understood it.  check later.}



The maximum of the density variance is at $\rho \approx 0.5, \epsilon
\approx 1$. This can be explained as follows: $\epsilon=1$ produces a
sharp separation of the jam (high density) and the laminar structure
(low density). With $\rho=0.5$ it turns out that these two phases have
the same length, thus \densVar\ is maximal. Increasing (decreasing)
$\rho$ will increase (decrease) the length of the jam and therefore in
both cases decreases \densVar.  \yy{why is this at $\rho=0.5$?}

The elliptical shape with its diagonal axis (instead of a vertial axis
as in Fig.~\ref{fig:RhoTemp}) can be explained as follows: Increasing
$\epsilon$ decreases, via Eq.~\ref{eq:SK_updateV}, the acceleration.
This leads to lower density in the laminar region.  Since $\densVar$
is maximal when both phases have the same length, and since a lower
density in the laminar region makes the jam occupy more space, one
needs to reduce the density in order to go back to the state where
they occupy equal space.

In summary, one obtains, for the traffic model, a phase diagram as in
Fig.~\ref{fig:RhoTemp}, which is the schematic phase diagram for a
gas-liquid transition in fluids.  Again, the important feature of this
phase diagram is that there are three states (laminar/gas;
mixed/coexistence; jammed/liquid) for low temperatures.  For higher
temperatures, the coexistence range becomes more and more narrow,
while the density of the gas phase and the density of the liquid phase
approach each other.  Eventually, they become equal, and the
coexistence state dies out.  The only important difference is that for
our traffic model the phase diagram is bent to the left with
increasing $\epsilon$.

There are other criteria which can be used to understand these types
of phase transitions.  In particular, one can look at the gap
distribution between jams, and one would expect a fractal structure at
the point where the 2-phase and the 1-phase model meet, i.e.\ at
$\rho \approx 0.2$ and $\epsilon \approx 1.7$.  This is still under
investigation but looks promising.  This should be followed by an
investigation in how far cell-based traffic models~\cite{Chow:etc:review} 
models display a similar transition.

\yy{need to relate to krauss model classes.  not clear where}

\yy{Note, not to be forgotten: AT the critical/fractal point, the
amount of noise should not matter.  Below that point, the amount of
noise probably will matter in the sense that it provides an upper
cut-off for the size of the condensation.}

\section{PHASE TRANSITIONS IN DETERMINISTIC MODELS}
\label{sec:det}

Only stochastic models can display spontaneous transitions between
homogeneous and coexistence ($=$ mixed) states.  The nature of the
transition can however also become clear in deterministic models.  We
will discuss these similarities first for a deterministic car
following model and then for deterministic fluid-dynamical models.

\subsection{Car Following Models}

For the model of \eq{eq:Bando}, it has been shown~\cite{Bando:etc:pre}
that the homogeneous solution of the model is linearly stable for all
densities for $V' < \alpha/2$, and linearly unstable for certain
densities for $V' > \alpha/2$, where $V'$ is the first derivative of
the function $V(.)$.  The instability sets in for certain intermediate
densities, that is, for low and high densities \emph{all} models are
stable in the homogeneous (laminar or jammed) state.  For intermediate
densities, one can select the curve $V(g)$ and the parameter
$\alpha$ such that the model is either stable or not stable.

If all parameters including the density are such that the homogeneous
solution is not stable, then the system rearranges itself into a
pattern of stop-and-go traffic, corresponding to the coexistence
state.  The density of the laminar and the jammed phase in the
coexistence state are independent from the average system density,
that is, if in that state system density goes up, it is reflected in
the jammed phase using up a larger fraction of space.

The type of the instability is similar to the better-known instability
of \eq{eq:theGang}.
However, once the instability is triggered in
\eq{eq:theGang}, it will just grow exponentially, and no stable
2-phase solution is found (e.g.~\cite{Gerlough:book}).

\subsection{Fluid-Dynamical Models}

Standard Lighthill-Whitham theory, of the type
\[
\partial_t \rho + \partial_x Q(\rho) = 0
\]
with a strictly convex flow-density-curve $Q(\rho)$, results in a
1-phase model.  When $Q(\rho)$ has linear sections, then in that range
shock waves are marginally stable, in the sense that disturbances to
those shocks are neither amplified nor dissipated away.  (In a 1-phase
model, disturbances are dissipated away and the final state is always
homogeneous; in a 2-phase/3-state model, there is a density regime
where disturbances to shock waves are dissipated away in the sense
that the shock wave will regain its standard profile again.)  When
$Q(\rho)$ has concave sections, then the situation becomes more
complicated (e.g.~\cite{Krug:Ferrari}).

Fluid-dynamical theory, of the type
\[
\partial_t \rho + \partial_x (\rho \, v) = 0
\]
and
\[
\partial_t v + v \, \partial_x v
= \frac{1}{\tau} \Big( V(\rho) - v \Big)
+ \alpha(\rho) \, \partial_x \rho
+ \nu(\rho) \, \partial_x^2 v
\]
\emph{can}, depending on the choice of parameters including the
$V(\rho)$-curve, either be a 1-phase/1-state or a 2-phase/3-state
model~\cite{Kuehne:93}.
\yy{can I look up the linear instab params?}

\bigskip

As pointed out before, these models are deterministic, so in
no situation will these models display stochastic transitions.

\section{DISCUSSION}

There is no general agreement if measurements show 1-phase/1-state or
2-phase/3-state traffic (or possibly even three
phases~\cite{Kerner:Rehborn:mea2}).  There is some evidence for
hysteresis in Germany~\cite{Kerner:Konh:cluster}, manifesting itself
in transitions from high to lower flow values at the same density.
Hysteresis, which was also found earlier~\cite{Treiterer:hysteresis},
is a strong indication for a 2-phase model.  However, even in Germany,
most measurements indicate highly variable traffic at intermediate
densities, which does not correspond to any clear-cut picture.

In this context, one should note that a 1-phase model which is close
to a 2-phase model would also display highly variable traffic at
intermediate densities, although it would be homogeneous at large
scales as discussed in Sec.~\ref{sec:phase-diag}.  This variability is
however a property of stochastic models only and for that reason it is
not well integrated into current theory development.  A precise
investigation of these relations is beyond the scope of this paper.
It seems however impossible to us to clarify the question if traffic
displays several phases or not -- and therefore, if breakdown
probability should be entered into the Highway Capacity Manual or not
-- without having understood how different phases are generated by
stochastic models.  The present paper fills exactly this gap.

\section{SUMMARY}

This paper shows, via numerical evidence, that a specific
stochastic car following model can either display 1-phase/1-state or
2-phase/3-state traffic, depending on the choice of parameters.  With
2-phase parameters, the two phases are laminar and jammed, which also
corresponds to two of the three states.  Those states are
homogeneous.  The third state, at intermediate densities, is a
coexistence or mixed state, consisting of sections with jammed and
sections with laminar traffic.  

The transition to a 1-phase/1-state model happens via the densities of
the laminar and of the jammed phase approaching each other until they
become the same.  Beyond this point, there is only one homogeneous
phase of traffic.  

Some of these findings can be understood by looking at deterministic
models for traffic, either car-following or fluid-dynamical.  However,
the stochastic elements of the transition cannot be explained by
deterministic models.  An important stochastic element is that
structure formation and strong variability can also happen in a
1-phase model as long as the parameters are close to the 2-phase
model -- a deterministic model would converge to a homogeneous
solution here. 

In our view, it is important to understand this possibility of
stochastic models to be in different regimes if one considers to enter
discussions of traffic breakdown probabilities into the Highway
Capacity Manual.  If traffic is best described by a 1-phase model,
then there is, in our view, no theoretical justification for such
probabilities.  If, however, traffic is best described by a 2-phase
model, then the 2-phase model could even give theoretical predictions
for breakdown probabilities.  A discussion of breakdown probabilities
in 2-phase models can be found in Ref.~\cite{Nagel:etc:tgf01}.

\section*{ACKNOWLEDGMENTS}

We thank F.~Wagner for intensive discussions.  Also D.~Stauffer and
M.~Droz have significantly helped in defining good computational
criteria for the detection of multiple phases.  P.~Wagner and
C.~Daganzo have helped indirectly (and maybe inadvertently), via the
discussion of a separate but related paper.  A significant portion of
the computing times on 192-CPU cluster Xibalba and on 44-CPU cluster
Linneus was used for some of the computational results.

\bibliographystyle{unsrtnat}
\bibliography{ref,kai,xref}
\end{document}